\documentclass{ws-procs9x6}
\pdfoutput=1
\usepackage{graphicx,xspace,units} 
\usepackage{float} 
\usepackage{amsmath} 
\usepackage{amssymb} 
\usepackage[normalem]{ulem} 
\usepackage{wrapfig} 
\usepackage{epsfig} 
\usepackage[usenames]{color} 
 
\definecolor{MidnightBlue}{cmyk}{0.98,0.13,0,0.43}
\definecolor{DarkGreen}{rgb}{0,0.7,0.1}

\newcommand{\be}{\begin{equation}} 
\newcommand{\ee}{\end{equation}}

\newcommand{\cE}{\mathcal{E}}

\newcommand{\curl}{\boldsymbol{\nabla} \times} 
 
\newcommand{\E}{\mathbf{E}} 
 
\newcommand{\B}{\mathbf{B}} 
 
\newcommand{\A}{\mathbf{A}} 
\newcommand{\J}{\mathbf{J}}

\newcommand{\tV}{\mathbb{V}} 
 
\newcommand{\tI}{\mathbb{I}}

\newcommand{\calE}{\mathcal{E}}

\newcommand{\dA}{\mathcal{D}\A}

\newcommand{\dJ}{\mathcal{D}\J} 
\newcommand{\dJJ}{\left. \dJ\dJ^* \right|_\text{obj}} 
\newcommand{\dJJprime}{\left. \dJ'{\dJ'}^{*} \right|_\text{obj}}

\newcommand{\tr}{\text{tr }}

\newcommand{\aindex}{\alpha} 
\newcommand{\bindex}{\beta} 
 
\newcommand{\abindex}{\aindex \bindex}

\newcommand{\F}{\mathbb{F}} 
\newcommand{\T}{\mathbb{T}}

\newcommand{\U}{\mathbb{U}} 
\newcommand{\V}{\mathbb{V}} 
 
\newcommand{\Hzero}{\mathbb{H}_0}

\newcommand{\Eout}{\E^\text{out}} 
\newcommand{\Ereg}{\E^\text{reg}} 
 
\newcommand{\Eregcc}{\E^{\text{reg}*}} 
\newcommand{\EregbQcc}{\Eregcc_{\bindex}} 
\newcommand{\EregaP}{\Ereg_{\aindex}} 
\newcommand{\EoutaP}{\Eout_{\aindex}}

\newcommand{\EregbQ}{\Ereg_{\bindex}} 
\newcommand{\EoutbQ}{\Eout_{\bindex}}

\newcommand{\tG}{\mathbb{G}} 
\newcommand{\tGzero}{\tG_0}

\newcommand{\vecx}{\mathbf{x}}

\newcommand{\vecX}{\mathbf{X}}

\begin{document}

\title{Casimir Physics: Geometry, Shape and Material}

\author{T. Emig}

\address{ Institut f\"ur Theoretische Physik, Universit\"at zu
  K\"oln, Z\"ulpicher Strasse 77, \\50937 K\"oln, Germany}
\address{Laboratoire de Physique Th\'eorique et Mod\`eles
  Statistiques, CNRS UMR 8626, B\^at.~100, Universit\'e Paris-Sud, 91405
  Orsay cedex, France}

\begin{abstract}
  The properties of fluctuation induced interactions like van der
  Waals and Casimir-Lifshitz forces are of interest in a plethora of
  fields ranging from biophysics to nanotechnology. Here we describe a
  general approach to compute these interactions. It is based on a
  combination of methods from statistical physics and scattering
  theory. We showcase how it is exquisitely suited to analyze a
  variety of previously unexplored phenomena.  Examples are given to
  show how the interplay of geometry and material properties helps to
  understand and control these forces.
\end{abstract}


\bodymatter

\section{Introduction}
\label{sec:intro}

All material objects, even if charge neutral, support instantaneous
current fluctuations due to quantum and thermal fluctuations of their
charge distribution. The interaction that results from the
electromagnetic coupling of these currents on different objects is
usually called the Casimir force. Originally, this force has been
derived for two parallel perfect metal plates \cite{Casimir48-1} and
atoms \cite{Casimir48-2}, and generalized later to two infinite
dielectric half-spaces with planar and parallel surfaces
\cite{Lifshitz55,Lifshitz56,Lifshitz57,Dzyaloshinskii61}.  The
non-additivity of the Casimir force limits these results in their
applicability to objects at very short separation via the so-called
proximity force approximation which provides only an uncontrolled
approximation of surface curvature to lowest order at vanishingly
small separations and ignores the global geometrical arrangement of
the objects. Generically, one encounters in practice geometries and
shapes that are rather distinct from infinite, parallel and planar
surfaces. Hence one faces the problem to compute the Casimir force
between objects of general shape, arrangement and material
decomposition. 

This article summarizes recent progress that has been proofed useful
in solving this problem for a variety of geometries.  (For an overview
of the development of related approaches, see Ref. \refcite{Rahi:fk}.)
In order to study Casimir forces in more general geometries, it turns
out to be advantageous to describe how fluctuating currents are
induced on the objects by the scattering of electromagnetic waves.
This representation of the Casimir interaction was developed in
Refs.~\refcite{Emig07,Emig08,Rahi:fk}. Each object is characterized by
its on-shell electromagnetic scattering amplitude.  The separations
and orientations of the objects are encoded in universal translation
matrices, which describe how a solution to the source-free Maxwell's
equations in the basis appropriate to one object looks when expanded
in the basis appropriate to another. These matrices hence describe the
electrodynamic interaction of the multipole moments associated with
the currents and depend on the displacement and orientation of
coordinate systems, but not on the shape and material of the objects
themselves.  The scattering amplitudes and translation matrices are
then combined in a simple formula that allows efficient numerical and,
in some cases, analytical calculations of Casimir forces and torques
for a wide variety of geometries, materials, and external conditions.
The approach applies to any finite number of arbitrarily shaped
objects with arbitrary linear electromagnetic response at zero or
finite temperature.

To illustrate this general formulation, we provide some sample
applications, including results for the interaction between metallic
objects for two spheres and for a sphere and a plane, taking into
account the combined effect of shape and material properties at large
distances.  In addition, we provide examples for the non-additivity of
the interaction by considering three objects (two spheres and a plane)
and for the orientation dependence in the case of spheroids. The
results are presented in form of analytical expressions at large
distances and as numerical results at smaller separations.

\section{Fluctuating currents and T-operators}
\label{sec:T-op}

We consider the Casimir energy for neutral objects with electric and
magnetic susceptibilities. The partition function $Z$ is defined
through the path integral, which sums all configurations of the
electromagnetic field (outside and inside the objects) with periodic
boundary conditions in time between $0$ and $T$.  The free energy $F$
of the field at inverse temperature $\beta$ is
\begin{equation}
F(\beta) = -\frac{1}{\beta}\log Z(\beta). 
\label{free} 
\end{equation}
The unrenormalized free energy generally depends on the ultraviolet
cutoff, but cutoff-dependent contributions arise from the objects
individually and do not depend on their separations or orientations.
Since we are only interested in energy differences, we can remove
these divergences by subtracting the energy of the system
when the objects are in some reference configuration, see below.  By
replacing the time $T$ by $-i\hbar\beta$, we obtain the partition
function $Z(\beta)$ in 4D Euclidean space.  In $A^0=0$ gauge, the
result is simply to replace the Matsubara frequencies $\omega_n =
\frac{2\pi n}{T}$ by $i\frac{2\pi n}{\hbar \beta}=ic\kappa_n$, where
$\kappa_n$ is the $n^{\rm th}$ Matsubara frequency divided by $c$.
The action is quadratic, so the modes with different $\kappa_n$
decouple and the partition function decomposes into a product of
partition functions for each mode. In the limit 
$\beta\to\infty$, the sum $\sum_{n\geq 0}$ turns into an integral  
$\frac{\hbar c \beta}{2\pi}\int_{0}^\infty d\kappa$, and we have 
the ground state energy
\begin{equation}
\calE_0 = -\frac{\hbar c}{2\pi} \int_0^\infty d\kappa \, 
\log Z(\kappa), 
\label{EKem} 
\end{equation}
with 
\begin{equation}
\begin{split} 
Z(\kappa) = \int \dA \dA^* \,  
\exp & \left[ -\beta \int d\vecx \, 
\E^{*}(\kappa,\vecx) \left(\Hzero+\frac{1}{\kappa^2} 
\tV(\kappa,\vecx) \right) \, \E (\kappa,\vecx)  
 \right], 
\end{split} 
\label{ZKem} 
\end{equation}
where we have used $\curl \E = i\frac{\omega}{c} \B$ to eliminate $\B$
in the action, and it is assumed that $\E$ is expressed by $\E= -
c^{-1} \partial_t \A$ in terms of the vector potential $\A$.  This
functional integral sums over configurations of $\A$. This sum
must be restricted by a choice of gauge, so that it does not include
the infinitely redundant gauge orbits. We will choose to work in the
gauge $A^0=0$, although of course no physical results depend on this
choice. Here we defined the Helmholtz operator
\begin{equation}
\Hzero(\kappa)=\tI +\frac{1}{\kappa^2} \nabla \times \nabla \times \, ,
\end{equation}
which is inverted by the
Green's function that is defined by
\begin{equation}
\kappa^2 \Hzero(\kappa) \tG_0(\kappa,\vecx,\vecx') = \tI
\delta^{(3)}(\vecx-\vecx') \, .
\end{equation}
The potential operator is
\begin{equation}
\tV(\kappa,\vecx) = \tI \, \kappa^2 
\left(\epsilon(ic\kappa,\vecx)-1\right) + \curl 
\left(\frac{1}{\mu(ic\kappa,\vecx)} -1 \right) \curl 
\,. 
\end{equation}
It is nonzero only at those points in space where the objects are
located ($\epsilon \neq 1$ or $\mu \neq 1$). At small frequencies,
typical materials have $\epsilon>1$ and $\mu\approx 1$, and $\V$ can
be regarded as an attractive potential.

Next, we transform to a free field (with kernel $\Hzero$) by
introducing fluctuating currents $\J$ that are confined to the
objects.  To perform this Hubbard-Stratonovich-like transformation we
multiply and divide the partition function of Eq.~(\ref{ZKem}) by
\begin{equation}
\begin{split} 
W & = \int \dJJ \exp\left[-\beta
\int d\vecx \, \J^*(\vecx) \cdot \V^{-1}(\kappa,\vecx) 
\J(\vecx) \right]  = \det \V \, , 
\end{split} 
\end{equation}
where $\left.\right|_{\rm obj}$ indicates that the currents are
defined only over the objects, {\it i.e.\/} the domain where $\V$ is
nonzero (and therefore $\V^{-1}$ exists), and we have represented the
local potential as a matrix in position space, $\V(\kappa,\vecx,
\vecx') = \V(\kappa,\vecx) \delta^{(3)}(\vecx - \vecx')$.  We then
change variables in the integration, $\J(\vecx) = \J'(\vecx) +
\frac{i}{\kappa} \V(\kappa,\vecx) \E(\vecx)$ and $\J^{*}(\vecx) =
{\J'}^{*}(\vecx) + \frac{i}{\kappa} \V(\kappa,\vecx) \E^{*}(\vecx)$,
to obtain
\begin{equation}
\label{EtoJ} 
\begin{split} 
\raisetag{20pt}
Z(\kappa) & = \frac{1}{W} \!\int\! \dA \dA^* \dJJprime \, 
\times\\ &
\exp \left[ 
-\beta \!\!\int \!\!d\vecx \, 
\E^{*}(\kappa,\vecx)
\left(\Hzero(\kappa)+
\frac{1}{\kappa^2} \V(\kappa,\vecx)\right)\E(\kappa,\vecx)\right.
\\ & \left. 
\!\!+ \left({\J'}^{*}(\vecx) + \frac{i}{\kappa}  
\V(\kappa,\vecx) \E^{*}(\kappa,\vecx)\right) 
\V^{-1}(\kappa,\vecx) 
\left(\J'(\vecx) + \frac{i}{\kappa}  
\V(\kappa,\vecx) \E(\kappa,\vecx)\right) 
\right], 
\\ & = \frac{1}{W} \!\int\! \dA \dA^* \dJJprime \,
\times\\ &
\exp \left[ 
-\beta \!\!\int \!\!d\vecx \, 
\E^{*} \Hzero \E + {\J'}^{*} \V^{-1}\J' +\frac{i}{\kappa} \left(
{\J'}^{*}\E + \J'\E^{*}\right)\right] \, .
\end{split} 
\end{equation}
Now the free electromagnetic field  can be integrated out
using $\Hzero^{-1} = \kappa^2\tG_0$, yielding
\begin{equation}
\begin{split} 
\raisetag{45pt}
Z(\kappa) & = \frac{Z_0}{W} \int 
\dJJprime \cr 
& \exp\left[ 
-\beta \!\!\int \!\!d\vecx d\vecx' \, 
{\J'}^{*}(\vecx) \left( 
\tGzero(\kappa,\vecx,\vecx') 
 + \V^{-1}(\kappa,\vecx)\delta^{3}(\vecx-\vecx')\right) \J'(\vecx') 
\right], 
\end{split} 
\label{Zemfull} 
\end{equation}
with $ Z_0 = \int \dA \dA^* \exp[-\beta \int d\vecx \, \E^*
\Hzero(\kappa)\E]$. Both factors $W$ and $Z_0$ contain
cutoff-dependent contributions but are independent of the separation
of the objects. Hence these factors cancel and can be ignored when we
consider a {\it change} in the energy due to a change of the object's
separations with the shape and the material composition of the objects
fixed.  The kernel of the action in Eq.~\eqref{Zemfull} is the inverse
of the T-operator, i.e., $\T^{-1} = \tGzero + \V^{-1}$ which is
equivalent to
\begin{equation}
  \label{eq:T-operator-def}
  \T = \V(\tI + \tGzero\V)^{-1} \, .
\end{equation}
The Casimir energy at zero temperature (without the cutoff-dependent
parts) is hence
\begin{equation}
  \label{eq:Energy_full}
  \calE = -\frac{\hbar c}{2\pi} \int_0^\infty d\kappa \log\det \T \, .
\end{equation}
The determinant is here taken over the spatial indices $\vecx$ and
$\vecx'$, which are restricted to the objects since $\T$ vanishes if
$\vecx$ or $\vecx'$ are not on an object. To compute the determinant
we start from the expression for $\T^{-1}$ which yields the reciprocal
of the determinant. We decompose $\T^{-1}$ by introducing separate
position space basis functions for each object. The projection of the
currents onto this basis defines the object's multipole moments. This
yields a division of $\T^{-1}$ into blocks where each block is labeled by an
object.  

\begin{figure}
\begin{center}
\includegraphics[width=0.65\linewidth]{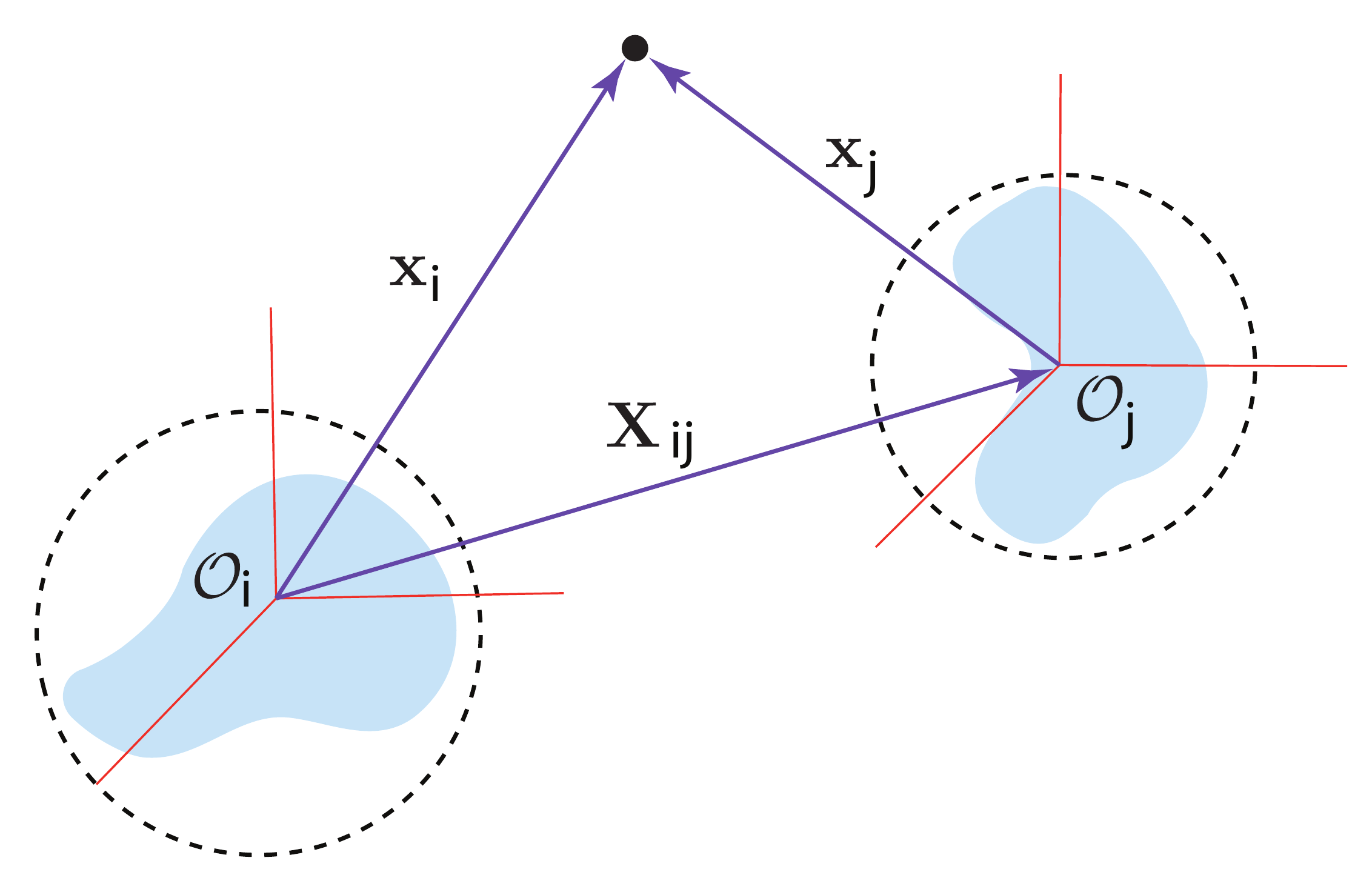} 
\end{center}
\caption{ Geometry of the configuration.  The dotted lines show
  surfaces separating the objects on which the radial variable is
  constant.  The translation vector $\vecX_{ij} = \vecx_i - \vecx_j =
  -\vecX_{ji}$ describes the relative positions of the two origins.  }
\label{fig:transboth} 
\end{figure}

The off-diagonal blocks are given by $\tGzero$ only and
describe the interaction of the multipoles on different objects. To
see this we choose for each object individually an eigenfunction basis
to expand the free Green's function, 
\begin{equation}
\label{G0-expansion}
\tGzero(\kappa,\vecx,\vecx')
= \sum_{\aindex} \EoutaP(\kappa,\vecx_>) \otimes \E^{\rm
  reg*}_\alpha(\kappa,\vecx'_<) 
\end{equation}
with regular solutions $\E^{\rm
  reg}_\alpha$ and outgoing solutions $\EoutaP$ of the free vector
Helmholtz equation, where $\vecx_<$ and $\vecx_>$ denote the position
with smaller and greater value of the ``radial'' variable of the
separable coordinates. The multipole moments of object $j$ are then
$Q_{j,\alpha}(\kappa)=\int d\vecx \J_j(\kappa,\vecx) \E^{\rm
  reg*}_\alpha(\kappa,\vecx) $. Regular solutions form a complete set
and hence outgoing solutions can be expanded in terms of regular
solutions except in a region (enclosed by a surface of constant radial
variable) that contains the origin of the
coordinate system of object $i$. This expansion defines the 
translation matrices $\U^{ji}_{\bindex,\aindex}$ via
\begin{equation}
\label{transoutreg} 
\EoutaP(\kappa,\vecx_i) = \sum_{\bindex} 
\U^{ji}_{\bindex\aindex}(\kappa,\vecX_{ji}) 
\EregbQ(\kappa,\vecx_j) \, ,
\end{equation}
where the definition of the coordinates is shown in
Fig.~\ref{fig:transboth}. The free Green's function then becomes
\begin{equation}
\tGzero(\kappa,\vecx,\vecx') =  \sum_{\aindex,\bindex} 
\EregaP(\kappa,\vecx_i) \otimes \U^{ji}_{\abindex}(\kappa,\vecX_{ji}) 
\EregbQcc(\kappa,\vecx_j') 
\end{equation}
so that the off-diagonal blocks of $\T^{-1}$ are given by the
translation matrices. Equivalent translation matrices can be defined
between two sets of regular solutions as is necessary  for one object
inside another, see Ref.~\refcite{Rahi:fk}. 

The diagonal blocks of $\T^{-1}$ are given by the matrix elements of
the T-operators $\T_j$ of the {\it individual} objects. By multiplying
$\T^{-1}$ by the T-operator $\T_\infty$ without the off-diagonal
blocks which can interpreted as describing a reference configuration
with infinite separations between the objects, one finds that (for
objects outside each other) the diagonal blocks are given by the
inverse of the matrix representing $\T_j$ in the basis $\EregaP$
\cite{Rahi:fk}. The physical meaning of this matrix follows from the
Lippmann-Schwinger equation for the full scattering solution
$\E_\alpha(\kappa,\vecx)$,
\begin{equation}
  \label{eq:lipp-schw}
  \E_\alpha(\kappa,\vecx)=\E^\text{reg}_\alpha(\kappa,\vecx) - \tGzero
  \V_j \E_\alpha(\kappa,\vecx) = \E^\text{reg}_\alpha(\kappa,\vecx) -
  \tGzero \T_j \E^\text{reg}_\alpha(\kappa,\vecx) \, .
\end{equation}
Using the expansion of Eq.~\eqref{G0-expansion}, the solution
sufficiently far away from the object (i.e., for positions that have
a radial variable larger than any point on the object) can be
expressed as
\begin{equation}
  \label{eq:scatt_amp}
  \E_\alpha(\kappa, \vecx)  =  \EregaP(\kappa, \vecx) -  
\sum_{\bindex} \EoutbQ(\kappa, \vecx) \int 
\EregbQcc(\kappa, \vecx') 
\T_j(\kappa) \EregaP(\kappa, \vecx') d\vecx'\, ,
\end{equation}
where the integral defines the scattering amplitude $\F_{j,\beta\alpha}(\kappa)$
of object $j$. It can be obtained, e.g., from matching boundary
conditions at the surface of a dielectric object. 

The Casimir energy (without cutoff-dependent contributions from $W$
and $Z_0$) can now be expressed as
\begin{equation}
  \label{Elogdet}
  \mathcal{E} = \frac{\hbar c}{2\pi} \int_0^\infty d\kappa  
\log \det (\mathbb{M} \mathbb{M}_\infty^{-1}), 
\end{equation}
where
\begin{equation}
  \mathbb{M} = 
\left( 
\begin{array}{c c c c} 
\F_1^{-1} & \U^{12} & \U^{13} & \cdots \\ 
\U^{21}    & \F_2^{-1} & \U^{23} & \cdots \\ 
\cdots & \cdots & \cdots & \cdots 
\end{array} 
\right) 
\end{equation}
and $\mathbb{M}^{-1}_\infty$ is the block diagonal matrix 
$\text{diag}(\F_1, \F_2, \cdots)$. For the case of two objects
this expressions simplifies to
\begin{equation}
  \label{Elogdet2}
  \mathcal{E} = \frac{\hbar c}{2\pi} \int_0^\infty d\kappa \log \det 
\left(\tI - \F_1\U^{12}\F_2\U^{21}\right) \, . 
\end{equation}
In order to obtain the free energy at nonzero temperature instead of 
the ground state energy, we do not take the limit $\beta\to\infty$ in 
Eq.~(\ref{free}) \cite{Lifshitz55}.  Instead, the integral $\frac{\hbar 
c}{2\pi} \int_0^\infty d\kappa$ is replaced everywhere by $\frac{1}{\beta} 
\sum_{n}'$, where $c \kappa_n=\frac{2\pi n}{\hbar\beta}$ with 
$n=0,1,2,3\ldots$ is the $n$th Matsubara frequency.  A careful 
analysis of the derivation shows that the zero frequency mode is 
weighted by $1/2$ compared to the rest of the terms in the sum; this 
modification of the sum is denoted by a prime on the summation 
symbol. 

\section{Applications}

In this section we demonstrate the applicability of the method through
some examples. Due to the lack of space, we only present the final
analytical and numerical results that all follow from
Eq.~\eqref{Elogdet} or Eq.~(\ref{Elogdet2}) by truncation of the
matrices at some order of partial waves, i.e., by considering only a
finite set of basis functions. At asymptotically large distances, the
interaction only depends on the dipole contribution while with
drecreasing distance the number of partial waves has to be increased.
Below we will provide results both in form of a asymptotic series in
the inverse separation and numerical results for a wide range of
distances.

\subsection{Sphere-plane}

First, we consider the sphere-plate geometry that has been
employed in the majority of recent experiments.  At large distances,
the energy can be expanded in an asymptotic series in the inverse 
separation. For a {\em dielectric sphere} in front
of {\em  perfectly reflecting mirror} with sphere-center to mirror separation
$L$ the Casimir  energy is 
\begin{equation}
  \label{eq:energy-eps-mu-sphere}
\begin{split}
\raisetag{46pt}
  {\mathcal E} &= -\frac{\hbar c}{\pi} \left\{
\frac{3}{8} (\alpha_1^\textsc{e} - \alpha_1^\textsc{m}) \frac{1}{L^4} 
+\frac{15}{32} (\alpha_2^\textsc{e} - \alpha_2^\textsc{m} +2 \gamma_{13}^\textsc{e} 
-2 \gamma_{13}^\textsc{m}) \frac{1}{L^6} \right.\\
& +\left. \frac{1}{1024} \left[ 23  (\alpha_1^\textsc{m})^2 - 14  
\alpha_1^\textsc{m}  \alpha_1^\textsc{e}
+23  (\alpha_1^\textsc{e})^2 
+ 2160 (\gamma_{14}^\textsc{e}- \gamma_{14}^\textsc{m})
\right] \frac{1}{L^7} \right. \\
& +\left.\frac{7}{7200} \left[ 572 (\alpha_3^\textsc{e}-\alpha_3^\textsc{m}) + 675 \left(
9(  \gamma_{15}^\textsc{e} -  \gamma_{15}^\textsc{m}) -55  
(  \gamma_{23}^\textsc{e} -  \gamma_{23}^\textsc{m})
\right)\right] \frac{1}{L^8} + \dots
\right\} \, ,
\end{split}
\end{equation}
where $\alpha_l^{\textsc{e}}$, $\alpha_l^{\textsc{m}}$ are the static
electric and magentic multipole polarizabilities of the sphere of
order $l$ ($l=2$ for dipoles), and the coefficients
$\gamma^{\textsc{e}}_{ln}$, $\gamma^{\textsc{m}}_{ln}$ describe
finite-frequency corrections to these polarizabilities, i.e., terms
$\sim \kappa^{2l+n}$ in the low-$\kappa$ expansion of the T-matrix
element for the $l^{\rm th}$ partial wave.  
Notice that the first three terms of the contribution at order $L^{-7}$ have
precisely the structure of the Casimir-Polder interaction between two
atoms with static dipole polarizabilities $\alpha_1^\textsc{m}$ and
$\alpha_1^\textsc{e}$ but it is reduced by a factor of $1/2^8$. This
factor and the distance dependence $\sim L^{-7}$ of this term suggests
that it arises from the interaction of the dipole fluctuations inside
the sphere with those inside its image at a distance $2L$. The
additional coefficient of $1/2$ in the reduction factor $(1/2)(1/2^7)$
can be traced back to the fact that the forces involved in bringing
the dipole in from infinity act only on the dipole and not on its
image. If the sphere is also assumed to be a {\em perfect reflector}, the energy
becomes
\begin{equation}
  \label{eq:EM-energy-series}
  {\mathcal E} = \frac{\hbar c}{\pi} \frac{1}{L} \sum_{j=4}^\infty b_j \left(
\frac{R}{L}\right)^{j-1} \, ,
\end{equation}
where the coefficients up to order $1/L^{11}$ are
\begin{eqnarray}
\label{eq:coeff_EM}
b_4&=&-\frac{9}{16}, \quad
b_5=0, \quad b_6=-\frac{25}{32}, \quad b_7=-\frac{3023}{4096} 
\nonumber\\ 
\quad b_8&=&-\frac{12551}{9600},
\quad  b_9=\frac{1282293}{163840},\nonumber \\
b_{10}&=&-\frac{32027856257}{722534400},  
\,\,\, b_{11}=\frac{39492614653}{412876800} \, .
\end{eqnarray}

Our method can be also employed to study the material
dependence of the interaction. When the sphere and the mirror
are described by a simple {\em plasma model}, we can obtain 
the interaction energy again from Eq.~\eqref{Elogdet2} by
substituting the dielectric function on the imaginary frequency
axis,
\begin{equation}
  \label{eq:epsilon_plasma}
    \epsilon_p(ic\kappa) = 1+
    \left(\frac{2\pi}{\lambda_p\kappa}\right)^2  \, ,
\end{equation}
into the T-matrices of sphere and mirror. From this we get at large
separations
\begin{equation}
  \label{eq:E_plasma}
    {\mathcal E} = -\frac{\hbar c}{\pi}\left[ f_4(\lambda_p/R) \frac{R^3}{L^4}
+f_5(\lambda_p/R)\frac{R^4}{L^5}  +{\cal O}(L^{-6}) \right] 
\end{equation}
with the functions
\begin{equation}
\label{eq:E_plasma_fcts}
\begin{split}
f_4(z) & =\frac{9}{16} +\frac{9}{64\pi^2} z^2 -\frac{9}{32\pi} z \coth\frac{2\pi}{z}\\
f_5(z) & = - \frac{13}{20\pi}z-\frac{21}{80\pi^3}
z^3+\frac{21}{40\pi^2} z^2 \coth \frac{2\pi}{z} \, .
\end{split}
\end{equation}
It is interesting that the amplitude $f_4$ of the leading term is not
universal but depends on the plasma wavelength $\lambda_p$. Only in
the two limits $\lambda_p/R\to 0$ and $\lambda_p/R\to\infty$ the 
amplitude assumes material independent values, $9/16$ and  $3/8$,
respectively. The first limit describes perfect reflection of electric
and magnetic fields at arbitrarily low frequencies and hence agrees
with the result of Eq.~\eqref{eq:EM-energy-series}.  The change to the
second amplitude for large $\lambda_p$ can be understood when one
considers a London superconductor that is described at zero
temperature by the plasma dielectric function \cite{Haakh:vn}. If one associates
$\lambda_p$ with the penetration depth, the perfect reflector limit
results from the absence of any field penetration while the second limit
corresponds to a large penetration depth and hence the suppression
of the magnetic mode contribution to the Casimir energy, explaining
the reduced amplitude of $3/8$. The latter result follows also when
the objects are considered to be normal metals, described by the
{\it Drude model} dielectric function
\begin{equation}
  \label{eq:eps_drude}
   \epsilon_p(ic\kappa)=1+\frac{(2\pi)^2}{(\lambda_p\kappa)^2+\pi c
     \kappa/\sigma} \, .
\end{equation}
From this function we get for a sphere and a mirror made of a Drude
metal the asymptotic energy
\begin{equation}
  \label{eq:e_drude}
   {\mathcal  E} = -\frac{\hbar c}{\pi} \left[
\frac{3}{8} \frac{R^3}{L^4} -\frac{77}{384} \frac{R^3}{\sqrt{2\sigma /c} \, L^{9/2}}
- \left(\frac{c}{8\pi \sigma}
-\frac{\pi}{20}\frac{\sigma R^2}{c}\right)\frac{R^3}{L^5}+{\cal
O}(L^{-\frac{11}{2}})\right] \, .
\end{equation}
In fact, one observes that the leading term is universal and agrees
with the $\lambda_p\to\infty$ limit of the plasma model. Note that the
result of Eq.~\eqref{eq:e_drude} does not apply to arbitrarily large
dc conductivity $\sigma$. The conditions for the validity of
Eq.~\eqref{eq:e_drude} can be written as $L\gg R$, $L\gg c/\sigma$ and
$L\gg \sigma R^2/c$. The above results demonstrate strong
correlations between shape and material since for two parallel,
infinite plates, both the plasma and the Drude model yield at large
separations the same (universal) result as a perfect mirror description.
 
\begin{figure}[h]
\begin{center}
\includegraphics[width=0.9\linewidth]{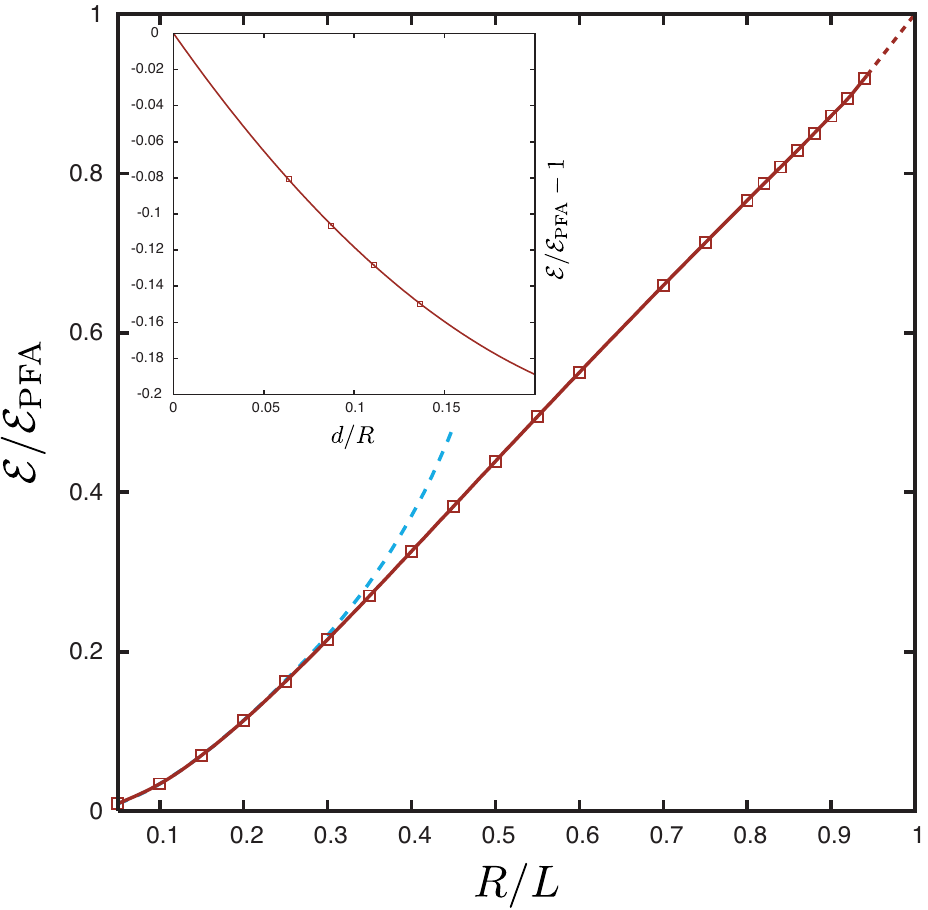}
\caption{Electromagnetic Casimir energy for the sphere-plate
  geometry. The energy is scaled by the proximity force approximation
  (PFA) energy ${\mathcal E}_\text{PFA} = -\frac{\pi^3}{720}
  \frac{\hbar c R}{d^2} $.  The asymptotic expansion of
  Eq.~(\ref{eq:EM-energy-series}) is shown as dashed line. Inset:
  Corrections to the PFA at small distances as function of $d=L-R$.}
\label{fig:EM}
\end{center}
\end{figure}

In order to study short separations, Eq.~\eqref{Elogdet2} has to be
evaluated numerically by including sufficiently many partial waves.
The result of an extrapolation from $l=29$ partial waves is shown in
Fig.~\ref{fig:EM} in the perfect reflection limit\cite{Emig08-1}. At
small separations the result can be fitted to a power law of the form
\begin{equation}
  \label{eq:E-PFA+corrections}
  {\mathcal E} = {\mathcal E}_\text{PFA} \left[ 1+ \theta_1 \frac{d}{R} +  \theta_2 
\left(\frac{d}{R}\right)^2 + \ldots \right] \, .
\end{equation}
with $ {\mathcal E}_\text{PFA} $ and $d$ defined in Fig. \ref{fig:EM}.
The coefficients $\theta_j$ measure corrections to the proximity force
approximation and are obtained from a fit of the function of
Eq.~(\ref{eq:E-PFA+corrections}) to the data points for the four
smallest studied separations. We find $\theta_1=-1.42 \pm 0.02$ and
$\theta_2=2.39 \pm 0.14$. This result is in agreement with numerical
findings in Ref.~\refcite{Maia_Neto08} but is in disagreement with an asymptotic
expansion for small distances \cite{Bordag:kx}. The latter yields
$\theta_1=-5.2$ and very small logarithmic corrections that however
can be ignored at the distances considered here. The origin of this
discrepancy is currently unclear but might be related to the
applicability of the asymptotic expansion to only much smaller
distances than accessible by current numerics.

\subsection{Three-body effects}

\begin{figure}[htbp]
\includegraphics[width=0.29\linewidth]{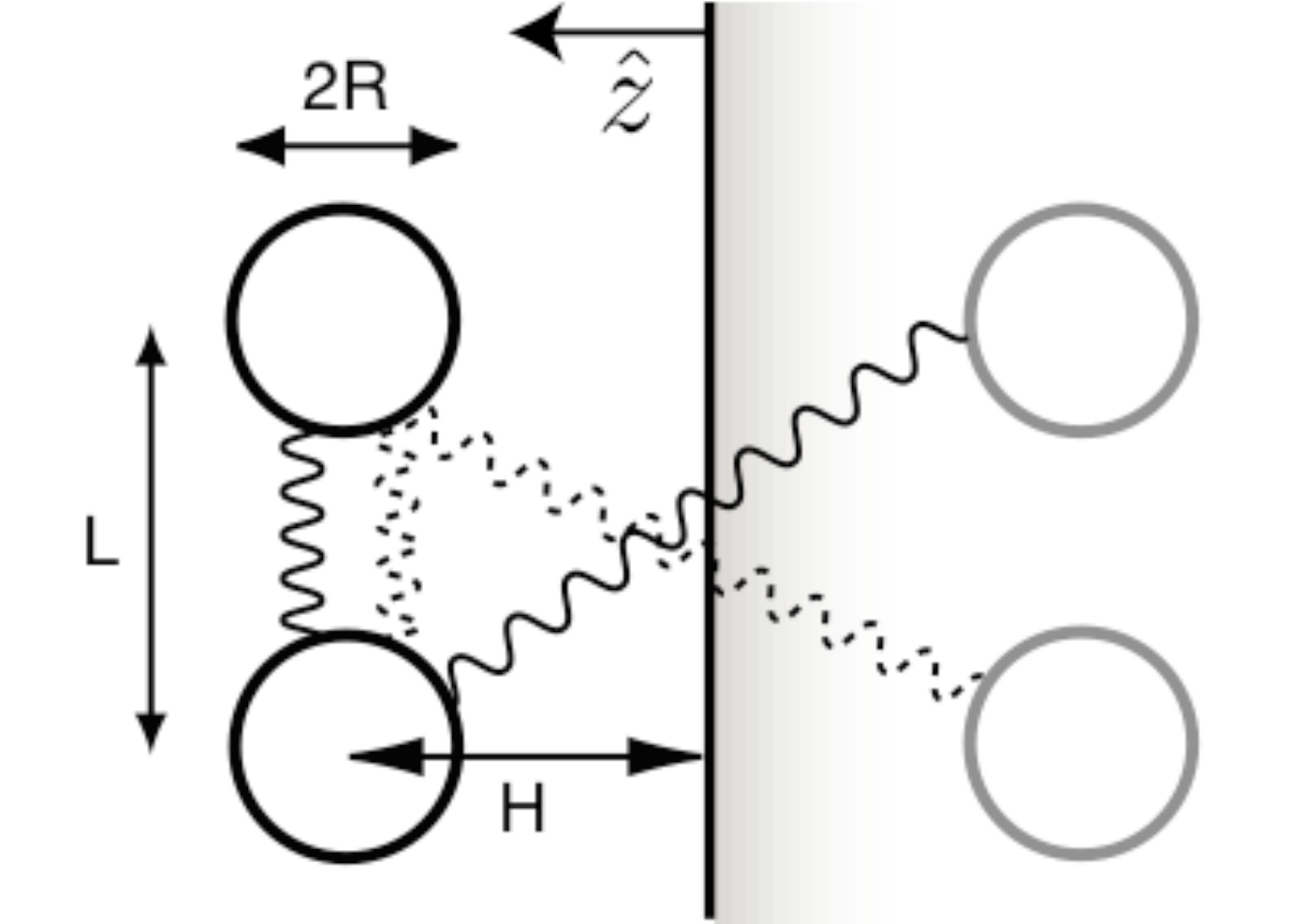}
\includegraphics[width=.7\linewidth]{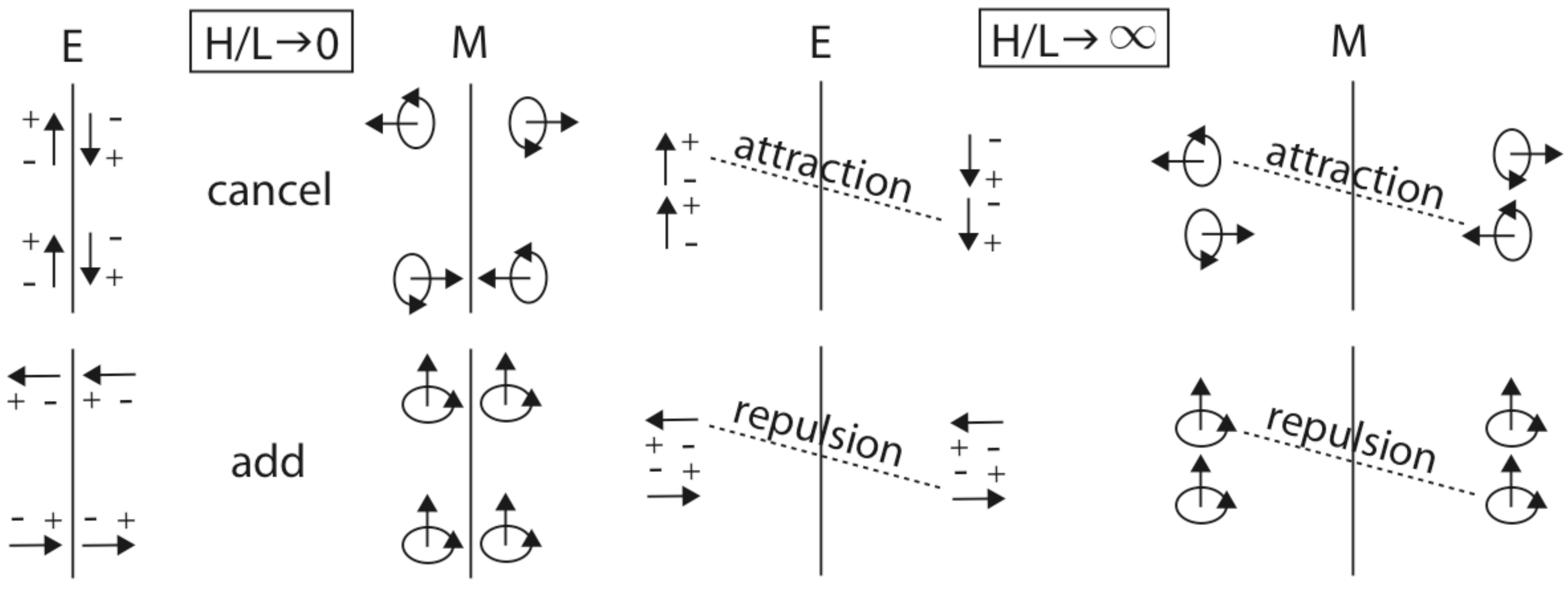}
\caption{\label{fig:2sphere+plate}Left: Geometry of the two-sphere/atom and
  sidewall system. Shown are also the mirror images (grey) and two-
  and three-body contributions (solid and dashed curly lines,
  respectively).  Right: Typical orientations of electric (E) and magnetic
  (M) dipoles and image dipoles for $H/L\to 0$ and $H/L\to\infty$.}
\end{figure}
Casimir interactions are not pair-wise additive. To study the
consequences of this property, we consider the case of two identical,
general polarizable objects near a perfectly reflecting wall in the
dipole approximation, see Fig.~\ref{fig:2sphere+plate}. This situation
applies to ground state atoms and also to general objects at {\it
  large} separations.  The separation between the objects is $L$ and
the separation of each of them from the wall is $H$.  In dipole
approximation, the retarded limit of the interaction is described by
the static electric ($\alpha_z$, $\alpha_\|$) and magnetic ($\beta_z$,
$\beta_\|$) dipole polarizabilities of the objects which can be
different in the directions perpendicular ($z$) and parallel ($\|$) to
the wall.  In the absence of the wall the potential for the  two
polarizable objects is given by  the well-known Casimir-Polder (CP) potential 
\begin{equation}
  \label{eq:E_CP}
{ \mathcal E}_{2,|}(L) = -\frac{\hbar c}{8\pi L^7} \!\!
\left[ 33 \alpha_\|^2 +\! 13 \alpha_z^2
- \! 14 \alpha_\|\beta_z + (\alpha \!\leftrightarrow\! \beta) \!\right] \, ,
\end{equation}
The $L$-dependent part of the interaction energy in the presence of
the wall is
\begin{equation}
  \label{eq:E_CP_plane}
  {\mathcal E}_{\underline{\circ\circ}}(L,H) = \cE_{2,|}(L) + \cE_{2,\backslash}(D,L) + \cE_3(D,L) 
\end{equation}
with $D=\sqrt{L^2+4H^2}$. The change in the relative orientation of the
objects with $\ell=L/D$ leads to the modified 2-body CP potential
\begin{equation}
\label{eq:E_CP_diag}
\begin{split}
\raisetag{35pt}
  \cE_{2,\backslash}(D,L) &= -\frac{\hbar c}{8\pi D^7} \!\!\left[ 26\alpha_\|^2
+\! 20 \alpha_z^2 -\! 14 \ell^2 (4\alpha_\|^2-9\alpha_\|\alpha_z
+5\alpha_z^2)\right.\\
& + \left. 63\ell^4 (\alpha_\| - \alpha_z)^2  
- 14\!\left(\alpha_\| \beta_\|(1\!-\!\ell^2) +\!\ell^2 \alpha_\| \beta_z \!\right) + (\alpha\!\leftrightarrow \!\beta) \right] \, .
\end{split}
\end{equation}
The 3-body energy $\cE_3(D,L)$ describes the collective interaction
between the two objects and one image object.  It is given by
\begin{equation}
  \label{eq:E_3}
\begin{split}
\raisetag{15pt}
  \cE_3(D,L) &=  \frac{4\hbar c}{\pi} \frac{1}{L^3D^4(\ell+1)^5}\left[ \Big(
3\ell^6 +15\ell^5+28\ell^4+20\ell^3+6\ell^2-5\ell-1\right)\\
&\times \left(\alpha_\|^2-\beta_\|^2\right)
- \left(3\ell^6+15\ell^5+24\ell^4-10\ell^2-5\ell-1\right) 
\left(\alpha_z^2-\beta_z^2\right)\\
& +4\left(\ell^4+5\ell^3+\ell^2\right)\left(\alpha_z\beta_\|-\alpha_\|\beta_z
\right)\Big] \, .
\end{split}
\end{equation}
It is instructive to consider the two limits $H\ll L$ and $H\gg L$.
For $H\ll L$ $\cE_{\underline{\circ\circ}}$ turns out to be the CP
potential of Eq.~\eqref{eq:E_CP} with the replacements $\alpha_z\to
2\alpha_z$, $\alpha_\|\to 0$, $\beta_z\to 0$, $\beta_\|\to
2\beta_\|$. The 2-body and 3-body contributions add constructively or
destructively, depending on the relative orientation of a dipole and
its image which together form a dipole of zero or twice the original
strength (see Fig.~\ref{fig:2sphere+plate}).

For $H \gg L$ the leading correction to the CP potential of
Eq.~\eqref{eq:E_CP} comes from the 3-body energy. The energy then
becomes (up to order $H^{-6}$)
\begin{equation}
  \label{eq:E_3_large_H}
   \cE_{\underline{\circ\circ}}(L,H) = { \mathcal E}_{2,|}(L)+\frac{\hbar c}{\pi} \!\!\left[ \!\frac{\alpha_z^2-\alpha_\|^2}{4 L^3H^4}  +
\frac{9\alpha_\|^2-\alpha_z^2 -2\alpha_\| \beta_z}{8LH^6} - (\alpha\leftrightarrow \beta)\!\right] .
\end{equation} 
The signs of the polarizabilities in the leading term $\sim H^{-4}$
can be understood from the relative orientation of the dipole of one
atom and the image dipole of the other atom, see Fig.~\ref{fig:2sphere+plate}.
If these two electric (magnetic) dipoles are almost perpendicular to
their distance vector they contribute attractively (repulsively) to
the potential between the two original objects. If these electric
(magnetic) dipoles are almost parallel to their distance vector they
yield a repulsive (attractive) contribution. For isotropic
polarizabilities the leading term of Eq.~\eqref{eq:E_3_large_H}
vanishes and the electric (magnetic) part $\sim H^{-6}$ of the 3-body
energy is always repulsive (attractive).

Next, we study the same geometry as before but with the objects assumed to
be two perfectly reflecting spheres of radius $R$.  The lengths $L$ and
$H$ are measured now from the centers of the spheres, see
Fig.~\ref{fig:2sphere+plate}. Here we do not limit the analysis to
large separations but consider arbitrary distances and include higher
order multipole moments than just dipole polarizability.  For $R \ll
L,\, H$ and arbitrary $H/L$ the result for the force can be written as
\begin{equation}
  \label{eq:force-of-L}
  F  = \frac{\hbar c}{\pi R^2} \sum_{j=6}^\infty  f_j(H/L) \left(\frac{R}{L}\right)^{j+2} \, .
\end{equation}
The functions $f_j$ can be computed exactly. We have obtained them up to $j=11$
and the first three are (with $s\equiv \sqrt{1+4h^2}$)
\begin{align}
  \label{eq:h-fcts}
& f_6(h) =  -\frac{1}{16h^8}\Big[s^{-9}(18 + 312 h^2 + 2052 h^4 + 6048 h^6 
\nonumber\\
&\! +  5719 h^8) + 18 - 12 h^2 + 1001 h^8\Big] \, , \quad f_7(h)=0\, , \\
& f_8(h) =  -\frac{1}{160 h^{12}}\Big[s^{-11} (6210 + 140554 h^2 + 1315364 h^4 
\nonumber\\
&\! + 6500242 h^6 + \! 17830560 h^8 + \! 25611168 h^{10} + \! 15000675 h^{12}) \nonumber\\
&\! - 6210 - 3934 h^2 + 764 h^4 - 78 h^6 + 71523 h^{12}\Big] \, .
\end{align}
For $H \gg L$ one has
$f_6(h) = -1001/16 +3/(4h^6)+ {\cal O}(h^{-8})$,
$f_8(h)=-71523/160+39/(80h^6)+ {\cal O}(h^{-8})$ so that the wall
induces weak repulsive corrections. For $H \ll L$,
$f_6(h)=-791/8+6741 h^2/8 +{\cal O}(h^4)$, $f_8(h)=-60939/80 + 582879
h^2/80 +{\cal O}(h^4)$ so that the force amplitude decreases  when the spheres are moved a small
distance  away from the wall. This proves the existence of a minimum in
the force amplitude as a function of $H/R$ for fixed, sufficiently
small $R/L$. We note that all $f_j(h)$ are finite for
$h\to \infty$ but some diverge for $h\to 0$, e.g., $f_9 \sim f_{11}
\sim h^{-3}$, making them important for small $H$.

\begin{figure}[htbp]
\begin{center}
\includegraphics[width=.8\linewidth]{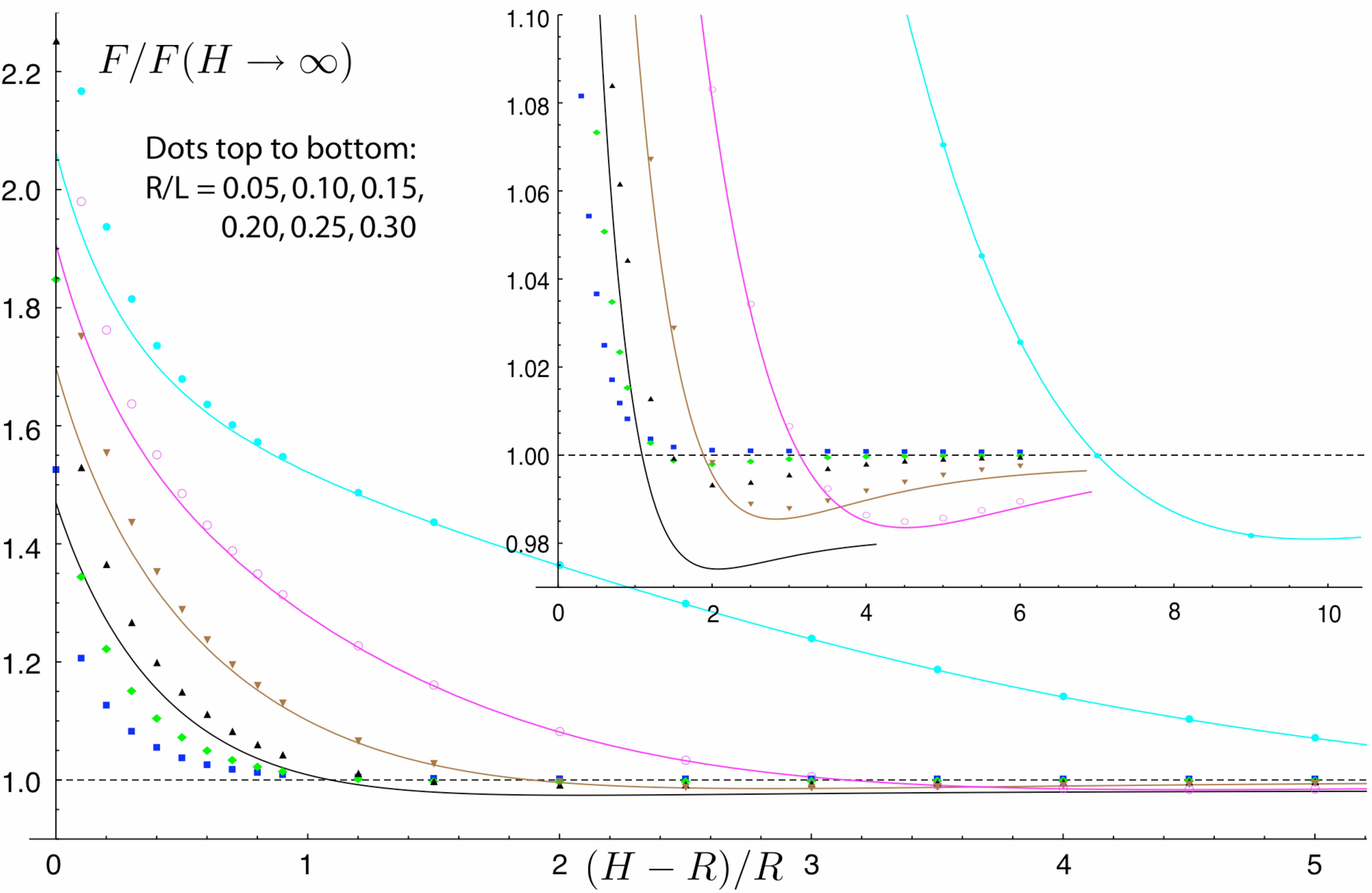}
\caption{\label{fig:2spheres+plate_num}Numerical results for the force (dots) between
  two spheres as function of the sidewall separation $H/R$ for
  different sphere separations $R/L$. Shown are also the analytical
  results of Eq.~\eqref{eq:force-of-L}, including terms up to $j=10$
  for $R/L\le 0.2$ (solid curves). Inset: Magnification of the
  nonmonotonicity.}
\end{center}
\end{figure}

To obtain the interaction at smaller separations or larger radius, we
have computed the energy $\cE_{\underline{\circ\circ}}$ and force
$F=-\partial \cE_{\underline{\circ\circ}} /\partial L$ between the
spheres numerically \cite{Rodriguez-Lopez:ys}.  In order to show the
effect of the wall, we plot the energy and force
normalized to the results for two spheres without a wall.
Fig.~\ref{fig:2spheres+plate_num} shows the force between the two
spheres as a function of the wall distance for fixed $L$. When the
spheres approach the wall, the force first decreases slightly if $R/L
\lesssim 0.3$ and then increases strongly under a further reduction of
$H$. For $R/L \gtrsim 0.3$ the force increases monotonically as the
spheres approach the wall.  This agrees with the prediction of the
large distance expansion. The expansion of Eq.~\eqref{eq:force-of-L}
with $j=10$ terms is also shown in Fig.~\ref{fig:2spheres+plate_num}
for $R/L\le 0.2$. Its validity is limited to large $L/R$ and not too
small $H/R$; it fails completely for $R/L>0.2$ and hence is not shown
in this range.

\subsection{Orientation dependence}

\begin{figure}[htbp]
\begin{center}
\includegraphics[width=.9\linewidth]{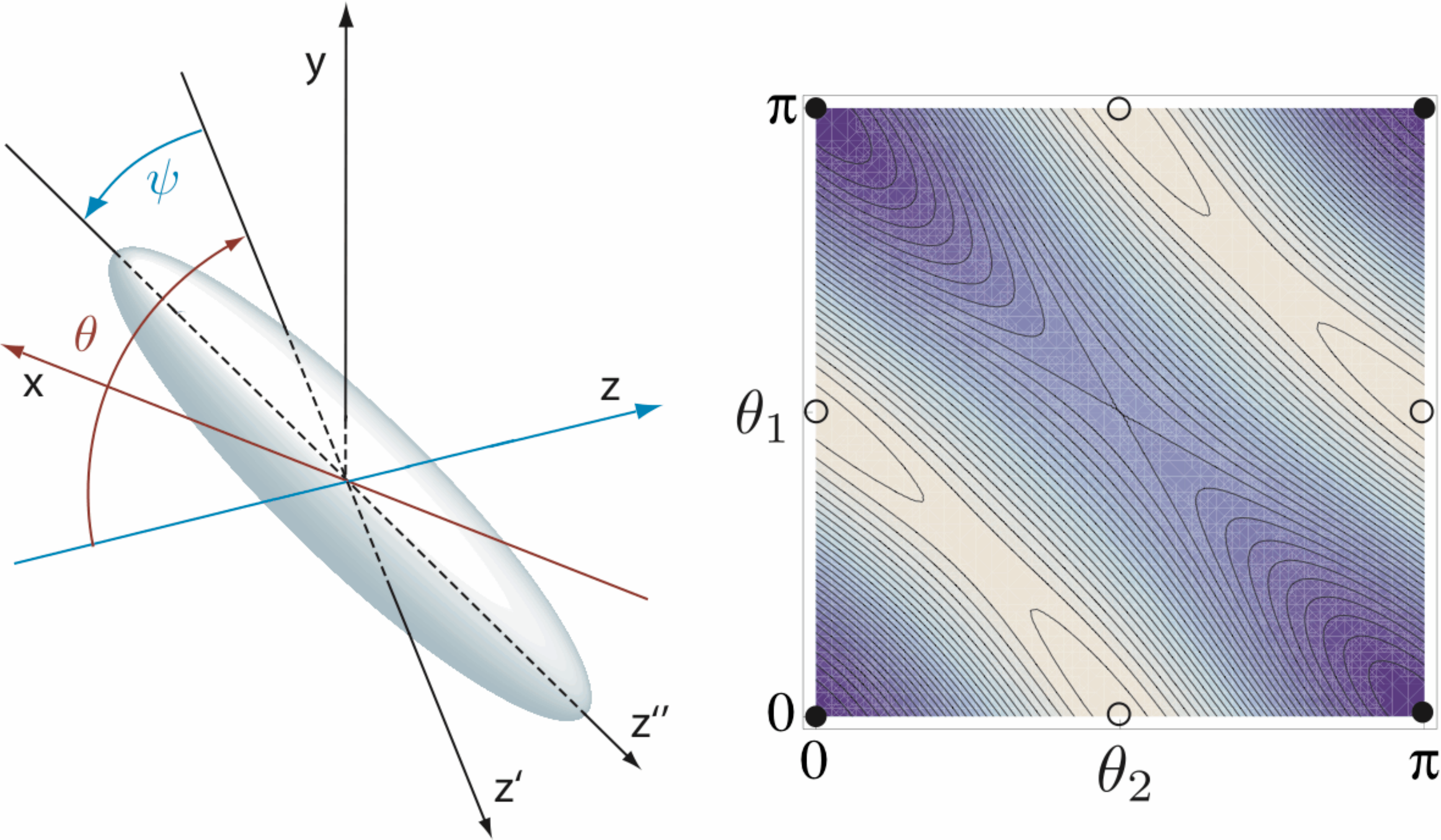}
\caption{\label{fig:cigars} (Color online) Orientation of a prolate (cigar-shaped)
  spheroid: The symmetry axis (initially the $z$-axis) is rotated by
  $\theta$ about the $x$-axis and then by $\psi$ about the $z$-axis.
  For two such spheroids, the energy at
  large distances is give by Eq.~\eqref{eq:energy-cylidenr-general}.
 The latter is depicted at fixed distance $d$, and for
  $\psi_1=\psi_2$, by a contour plot as function
  of the angles $\theta_1$, $\theta_2$ for the $x$-axis rotations . 
  Minima (maxima) are marked by filled (open) dots.}
\end{center}
\end{figure}

In this section we investigate the shape and orientation dependence of
the Casimir force using Eq.~\eqref{Elogdet2}. As examples we focus on
ellipsoids, computing the orientation dependent force between two
spheroids, and between a spheroid and a plane \cite{Emig:2009zr}.  For
two anisotropic objects, the CP potential of Eq.~\eqref{eq:E_CP} must
be generalized.  In terms of the Cartesian components of the standard
electric (magnetic) polarizability matrix $\mathbb{\alpha}$
($\mathbb{\beta}$), the asymptotic large distance potential of two
objects (with the $\hat{z}$ axis pointing from one object to the
other), can be written as
\begin{equation}
 \label{eq:energy_aniso}
\begin{split}
 \cE &=  -\frac{\hbar c}{d^7} \frac{1}{8\pi} \bigg\{
13\left( \alpha^1_{xx}\alpha^2_{xx} + \alpha^1_{yy}\alpha^2_{yy}+2 \alpha^1_{xy}\alpha^2_{xy}\right) \\
&+ 20 \, \alpha^1_{zz}\alpha^2_{zz} -30 \left( \alpha^1_{xz}\alpha^2_{xz} 
+ \alpha^1_{yz}\alpha^2_{yz}\right) +
\left(\mathbb{\alpha}\to\mathbb{\beta}\right) \\
&- 7 \left( \alpha^1_{xx}\beta^2_{yy} +  \alpha^1_{yy}\beta^2_{xx} 
-2 \alpha^1_{xy}\beta^2_{xy} \right) +\left( 1\leftrightarrow 2\right)
\bigg\} \, .
\end{split}
\end{equation} 
 For the case of an
ellipsoidal object with static electric permittivity $\epsilon$ and
magnetic permeability $\mu$, the polarizability tensors are diagonal
in a basis oriented to its principal axes, with elements (for
$i\in\{1,2,3\}$)
\begin{equation}
\label{eq:pol-tensor-diag}
\alpha_{ii}^0 = \frac{V}{4\pi} \frac{\epsilon-1}{1+(\epsilon-1)n_i}\, ,\,
\beta_{ii}^0 = \frac{V}{4\pi} \frac{\mu-1}{1+(\mu-1)n_i}\,,
\end{equation}
where $V=4\pi r_1 r_2 r_3/3$ is the ellipsoid's volume. In the case of
spheroids, for which $r_1=r_2=R$ and $r_3 = L/2$, the so-called
depolarizing factors can be expressed in terms of elementary
functions,
\begin{equation}
n_1=n_2=\frac{1-n_3}{2}, \, n_3 = \frac{1-e^2}{2e^3} \left(\log
\frac{1+e}{1-e} - 2 e \right),
\label{eq:depolarizing}
\end{equation}
where the eccentricity $e = \sqrt{1 - \frac{4R^2}{L^2}}$ is real for a
prolate spheroid ($L > 2R$) and imaginary for an oblate spheroid ($L <
2R$). The polarizability tensors for an arbitrary orientation are then
obtained as $\mathbb{\alpha}={\cal R}^{-1}\mathbb{\alpha}^0{\cal R}$,
where ${\cal R}$ is the matrix that rotates the principal axis of the
spheroid to the Cartesian basis, i.e.  ${\cal
  R}(1,2,3)\to(x,y,z)$. Note that for rarefied media with
$\epsilon\simeq 1$, $\mu\simeq 1$ the polarizabilities are isotropic
and proportional to the volume.  Hence, to leading order in
$\epsilon-1$ the interaction is orientation independent at
asymptotically large separations, as we would expect, since pairwise
summation is valid for $\epsilon-1\ll 1$. In the following we focus on
the interesting opposite limit of two identical perfectly reflecting
spheroids. We first consider prolate spheroids with $L \gg R$.  The
orientation of each ``needle'' relative to the line joining them (the
initial $z$-axis) is parameterized by the two angles $(\theta,\psi)$,
as depicted in Fig.~\ref{fig:cigars}.  Then the energy is
\begin{equation}
\label{eq:energy-cylidenr-general}
\begin{split}
\raisetag{55pt}
{\cal E}(\theta_1,\theta_2,\psi) &= -\frac{\hbar c}{d^7} \bigg\{
\frac{5L^6}{1152 \pi \left( \ln \frac{L}{R} - 1\right)^2}
\bigg[\cos^2\theta_1 \cos^2\theta_2\\
+ &\frac{13}{20}\cos^2\psi \sin^2 \theta_1\sin^2\theta_2
- \frac{3}{8} \cos\psi \sin 2\theta_1 \sin 2\theta_2\bigg]
+{\cal O}\bigg(\frac{L^4R^2}{\ln\frac{L}{R}}\bigg)\bigg\}\, ,
\end{split}
\end{equation}
where $\psi\equiv\psi_1-\psi_2$.  It is minimized for two
needles aligned parallel to their separation vector.  At almost all
orientations the energy scales as $L^6$, and vanishes logarithmically
slowly as $R\to 0$.  The latter scaling changes when one needle is
orthogonal to $\hat{z}$ (i.e. $\theta_1=\pi/2$), while the other is
either parallel to $\hat{z}$ ($\theta_2=0$) or has an arbitrary
$\theta_2$ but differs by an angle $\pi/2$ in its rotation about the
$z$-axis (i.e. $\psi_1-\psi_2=\pi/2$).  In these cases the energy
comes from the next order term in
Eq.~(\ref{eq:energy-cylidenr-general}), and takes the form
\begin{equation}
  \label{eq:crossed-cigars-finite-theta}
  {\cal E}\left(\frac{\pi}{2},\theta_2,\frac{\pi}{2}\right) = 
-\frac{\hbar c}{1152 \pi \, d^7} \frac{L^4R^2}{\ln\frac{L}{R} - 1} 
\left( 73+7\cos 2\theta_2
  \right) \, ,
\end{equation}
which shows that the least favorable configuration corresponds to two
needles orthogonal to each other and to the line joining them.

\begin{figure}[htbp]
\begin{center}
\includegraphics[width=.9\linewidth]{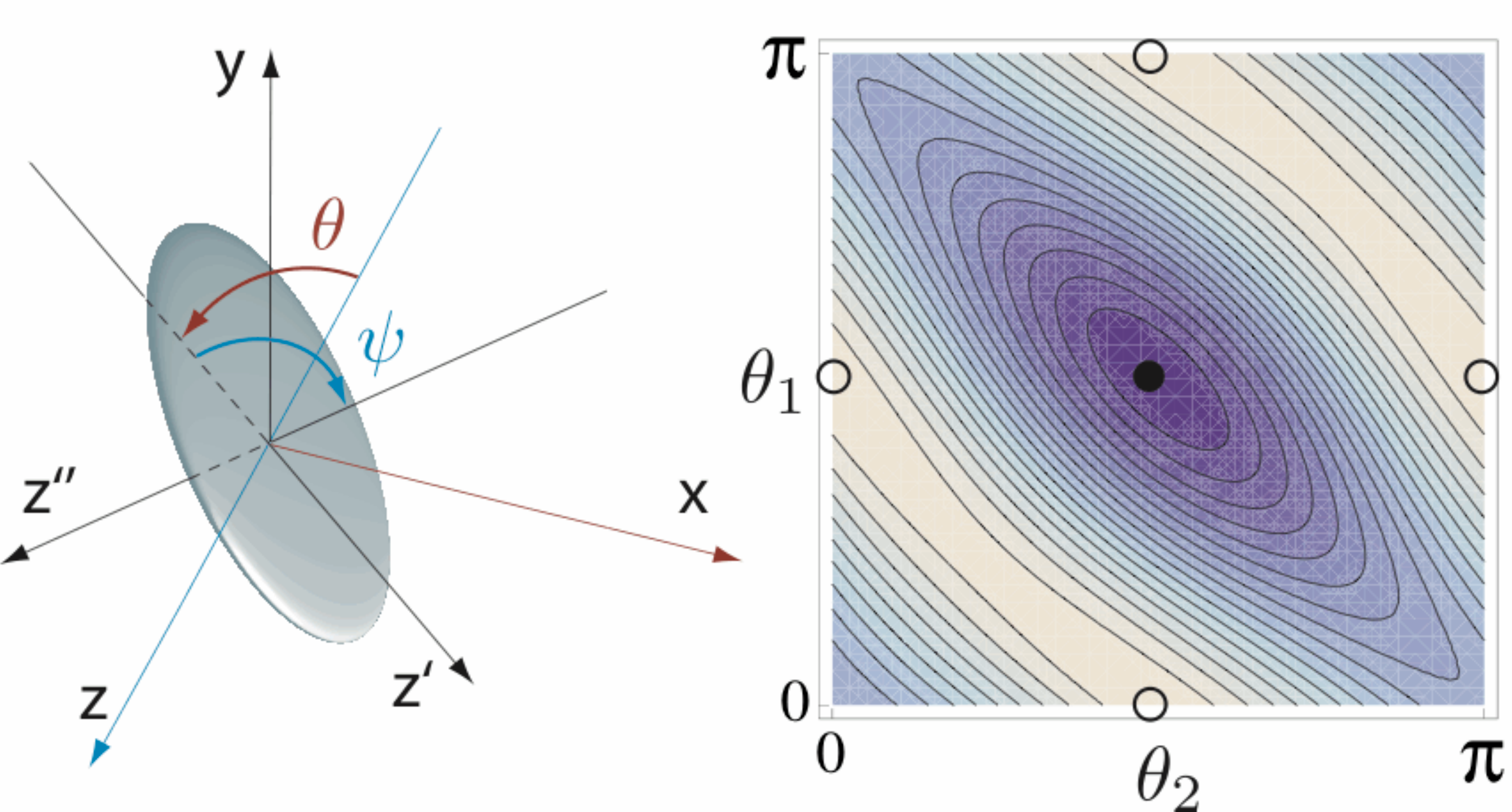}
\caption{\label{fig:pancakes} (Color online) As in
  Fig.~\ref{fig:cigars} for oblate (pancake-shaped) spheroids, with a
  contour plot of energy at large separations.}
\end{center}
\end{figure}

For perfectly reflecting oblate spheroids with $R\gg L/2$, the
orientation of each ``pancake'' is again described by a pair of angles
$(\theta,\psi)$, as depicted in Fig.~\ref{fig:pancakes}. To leading
order at large separations, the energy is given by
\begin{equation}
  \label{eq:energy_oblate}
\begin{split}
  \cE &= -\frac{\hbar c}{d^7} \bigg\{
\frac{R^6}{144\pi^3} \bigg[
765 - 5(\cos 2\theta_1+\cos 2\theta_2) 
+237 \cos 2\theta_1 \cos 2\theta_2 \\
&+372 \cos 2\psi \sin^2\theta_1\sin^2\theta_2 
- 300 \cos \psi\sin 2\theta_1 \sin 2\theta_2 \bigg] 
+{\cal O}\big( {R^5L}\big)\bigg\} \, .
\end{split}
\end{equation}
The leading dependence is proportional to $R^6$, and does not
disappear for any choice of orientations.  Furthermore, this
dependence remains even as the thickness of the pancake is taken to
zero ($L\to 0$). This is very different from the case of the needles,
where the interaction energy vanishes with thickness as
$\ln^{-1}(L/R)$.  The lack of $L$ dependence is due to the assumed
perfectly reflectivity.  The energy is minimal for two pancakes
lying on the same plane ($\theta_1=\theta_2=\pi/2$, $\psi=0$) and has
energy $-\hbar c \, (173/18\pi^3) R^6/d^7$.  When the two pancakes are
stacked on top of each other, the energy is increased to $-\hbar c
\,(62/9\pi^3) R^6/d^7$.  The least favorable configuration is when the
pancakes lie in perpendicular planes, i.e., $\theta_1=\pi/2$,
$\theta_2=0$, with an energy $-\hbar c\, (11/3\pi^3) R^6/d^7$.

For an anisotropic object interacting with a perfectly reflecting
mirror, at leading order the CP potential generalizes to
\begin{equation}
\label{eq:energy_aniso_wall}
\cE = -\frac{\hbar c}{d^4} \frac{1}{8\pi} \tr (\alpha-\beta ) 
+{\cal O}(d^{-5})\, ,
\end{equation}
which is clearly independent of orientation.  Orientation dependence
in this system thus comes from higher multipoles.  The next order also
vanishes, so the leading term is the contribution from the partial
waves with $l=3$ for which the scattering matrix is not known
analytically.  However, we can obtain the preferred orientation by
considering a distorted sphere in which the radius $R$ is deformed to
$R+\delta f(\vartheta,\varphi)$.  The function $f$ can be expanded
into spherical harmonics $Y_{lm}(\vartheta,\varphi)$, and spheroidal
symmetry can be mimicked by choosing $f=Y_{20}(\vartheta,\varphi)$.
The leading orientation dependent part of the energy is then obtained
as
\begin{equation}
\cE_f = - \hbar c \frac{1607}{640 \sqrt{5} \pi^{3/2}} \frac{\delta R^4}{d^6} \cos(2\theta)   \,. 
\end{equation}
A prolate spheroid ($\delta>0$) thus minimizes its energy by pointing
towards the mirror, while an oblate spheroid ($\delta<0$) prefers to
lie in a plane perpendicular to the mirror.  (We assume that the
perturbative results are not changed for large distortions.)  These
configurations are also preferred at small distances $d$, since (at
fixed distance to the center) the object reorients to minimize the
closest separation.  Interestingly, the latter conclusion is not
generally true. In Ref.~\refcite{Emig:2009zr} it has been shown that there
can be a transition in preferred orientation as a function of $d$ in
the simpler case of a scalar field with Neumann boundary conditions.
The separation at which this transition occurs varies with the
spheroid's eccentricity.

\subsection{Material dependence}

In this section we shall discuss some characteristic effects of the
Casimir interaction between metallic nano-particles by studying two spheres
with {\it finite} conductivity in the limit where their radius $R$ is
much smaller than their separation $d$. We assume further that $R$ is
large compared to the inverse Fermi wave vector $\pi/k_F$ of the
metal. Since typically $\pi/k_F$ is of the order of a few Angstrom,
this assumption is reasonable even for nano-particles.
Theories for the optical properties of small
metallic particles \cite{Wood:1982nl} suggest a Drude dielectric function
\begin{equation}
  \label{eq:eps_Drude}
  \epsilon(ic\kappa) = 1+4\pi \frac{\sigma(ic\kappa)}{c\kappa} \, ,
\end{equation}
where $\sigma(ic\kappa)$ is the conductivity which approaches for
$\kappa\to 0$ the dc conductivity $\sigma_{dc}$. For bulk metals
$\sigma_{dc}=\omega_p^2\tau/4\pi$ where $\omega_p=\sqrt{4e^2k_F^3/3\pi
  m_e}$ is the plasma frequency with electron charge $e$ and electron
mass $m_e$, and $\tau$ is the relaxation time. With decreasing
dimension of the particle, $\sigma_{dc}(R)$ is reduced compared to its
bulk value due to finite size effects and hence becomes a function of
$R$ \cite{Wood:1982nl}. In analogy to the result for a sphere and a
plate that are described by the Drude model, we obtain for the large
distance expansion of the energy the result
\begin{equation}
  \label{eq:E_drude_spheres}
\cE = -\hbar c \, \frac{23}{4\pi} \frac{R^6}{L^7} -\left(
\frac{R\sigma_{dc}(R)}{c} -\frac{45}{4\pi^2} \frac{c}{R\sigma_{dc}(R)}\right) \frac{R^7}{L^8} + \ldots \, .   
\end{equation}
As in the sphere-plate case, the leading term is material independent
but different from that of the perfect metal limit (where the
amplitude is $143/16\pi$) since only the electric polarization
contributes. At next order, the first and second terms in the
parentheses come from magnetic and electric dipole fluctuations,
respectively. The term $\sim 1/L^8$ is absent in the interaction
between perfectly conducting spheres. The limit of perfect
conductivity, $\sigma_{dc}\to\infty$ cannot be taken in
Eq.~(\ref{eq:E_drude_spheres}) since this limit does not commute with
the large $L$ expansion.

\begin{figure}[h]
\includegraphics[width=11cm]{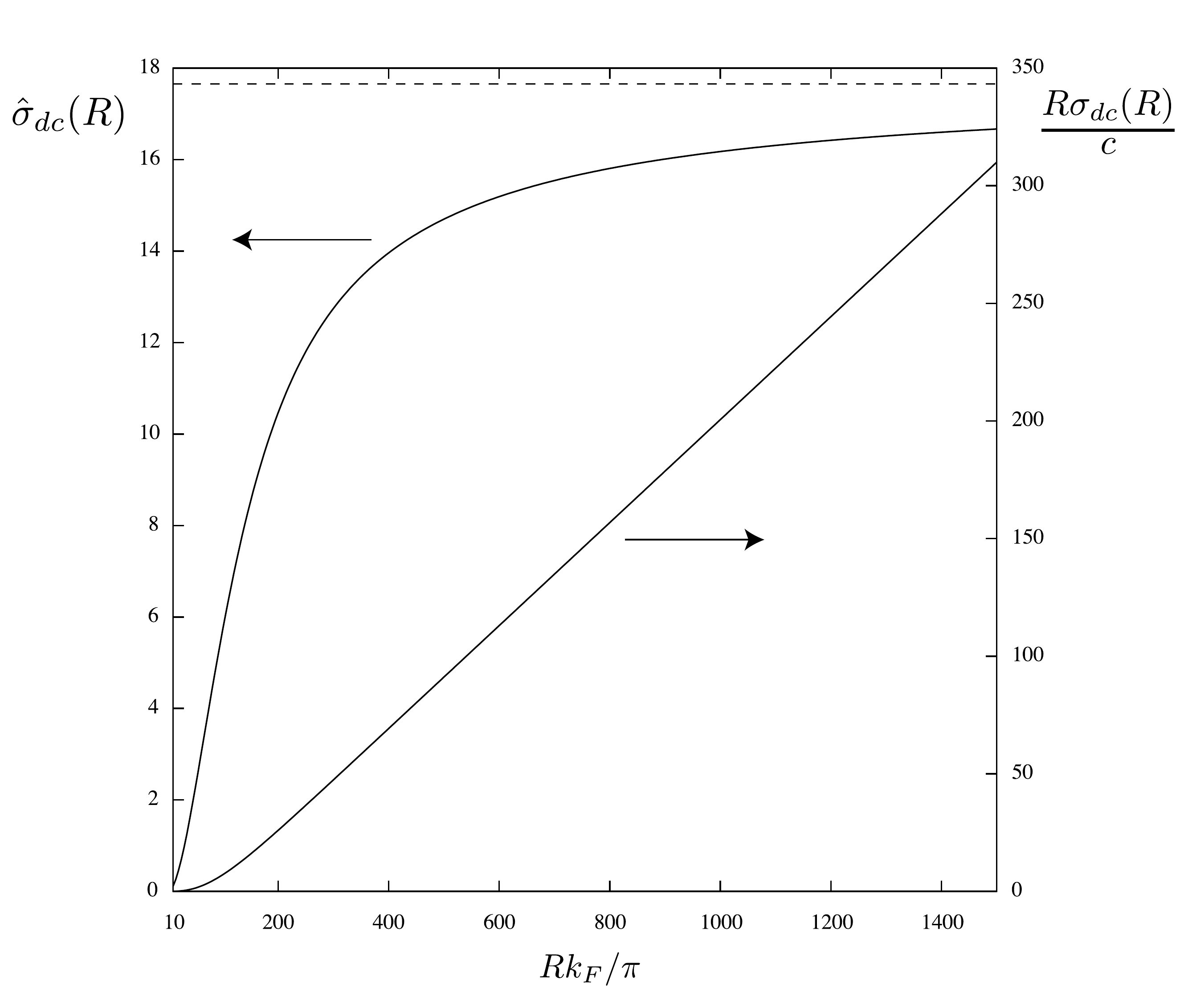}
\caption{\label{Fig:sigma} Dimensionless dc conductivity
  $\hat\sigma_{dc}(R)$ in units of $e^2/2\hbar a_0$ (with Bohr radius
  $a_0$) for a Aluminum sphere with $\epsilon_F=11.63$eV,
  $\pi/k_F=1.8\!\!\buildrel _\circ \over {\mathrm{A}}$ and
  $\tau=0.8\cdot 10^{-14}$sec as function of the radius $R$, measured
  in units of $\pi/k_F$. Also shown is the corresponding ratio
  $R\sigma_{dc}(R)/c$ that determines the Casimir interaction of
  Eq.~(\ref{eq:E_drude_spheres}). The bulk dc conductivity
  $\hat\sigma_{dc}(\infty)=17.66$ is indicated by the dashed line.}
\end{figure}

In order to estimate the effect of finite conductivity and its
dependence on the size of the nano-particle, we have to employ a
theory that can describe the evolution of $\sigma_{dc}(R)$ with the
particle size. A theory for the dielectric function of a cubical
metallic particle of dimensions $R \gg \pi/k_F$ has been developed
within the random phase approximation in the limit of low frequencies
$\ll c/R$ \cite{Wood:1982nl}. In this theory it is further assumed
that the discreteness of the electronic energy levels, and not the
inhomogeneity of the charge distribution, is important. This implies
that the particle responds only at the wave vector of the incident
field which is a rather common approximation for small particles.
From an electron number-conserving relaxation time approximation the
complex dielectric function is obtained which yields the
size-dependent dc conductivity for a cubic particle of volume $a^3$
\cite{Wood:1982nl}. It has been shown that the detailed shape of the
particle does not matter much, and we can set $a=(4\pi/3)^{1/3}R$
which defines the volume equivalent sphere radius $R$.  For
$\pi/k_F\simeq a$ the nano particle ceases to be conducting,
corresponding to a metal-insulator transition due to the localisation
of electrons for particles with a size of the order of the mean free
path.  It is instructive to consider the size dependence of
$\sigma_{dc}(R)$ and of the Casimir interaction for a particular
choice of material. Following Ref.~\refcite{Wood:1982nl}, we focus on
small Aluminum spheres with Fermi energy $\epsilon_F=11.63$eV and
$\tau=0.8\cdot 10^{-14}$sec. These parameters correspond to
$\pi/k_F=1.8\!\!\buildrel _\circ \over {\mathrm{A}}$ and a plasma
wavelength $\lambda_p=79$nm. It is useful to introduce the
dimensionless conductivity $\hat\sigma_{dc}(R)$, which is measured in
units of $e^2/2\hbar a_0$ with Bohr radius $a_0$, so that the
important quantity of Eq.~(\ref{eq:E_drude_spheres}) can be written as
$R\sigma_{dc}(R)/c=(\alpha/2)(R/a_0)\hat\sigma_{dc}(R)$ where $\alpha$
is the fine-structure constant. The result is shown in
Fig.~\ref{Fig:sigma}.  For example, for a sphere of radius $R=10$nm,
the dc conductivity is reduced by a factor $\approx 0.15$ compared to
the bulk Drude value.  If the radius of the sphere is equal to the
plasma wavelength $\lambda_p$, the reduction factor $\approx
0.8$. These results show that shape and material properties are
important for the Casimir interaction between
nano-particles. Potential applications include the interaction between
dilute suspensions of metallic nano-particles.

\subsection{Further extensions}

The general result of Eq.~\eqref{Elogdet} and its extensions described
in Ref.~\refcite{Rahi:fk} have been recently applied to a number of
new geometries and further applications are under way. Examples
include so-called interior configurations with an object contained
within an otherwise empty, perfectly conducting spherical shell
\cite{Zaheer:ve}. For this geometry the forces and torques on a
dielectric or conducting object, well separated from the cavity walls,
have been determined. Corrections to the proximity force approximation
for this interior problem have been obtained by computing the
interaction energy of a finite-size metal sphere with the cavity walls
when the separation between their surfaces tends to zero.
Eq.~\eqref{Elogdet}, evaluated in parabolic cylinder coordinates, has
been used to obtain the interaction energy of a parabolic cylinder and
an infinite plate (both perfect mirrors), as a function of their
separation and inclination, and the cylinder's parabolic radius
\cite{Graham:2009ly}. By taking the limit of vanishing radius,
corresponding to a semi-infinite plate, the effect of edge and
inclination could be studied.

\section*{Acknowledgments}

The reported results have been obtained in collaboration with
N. Graham, R. L. Jaffe, M. Kardar, S. J. Rahi, P. Rodriguez-Lopez,
A. Shpunt, S.~Zaheer, R. Zandi.  This work was supported by the
Deutsche Forschungsgemeinschaft (DFG) through grant EM70/3 and Defense
Advanced Research Projects Agency (DARPA) contract No. S-000354.

\end{document}